\documentstyle[12pt]{article}
\centerline{\bf{\Large
Magnetic properties of confined bosonic}}
\centerline{\bf {\Large vacuum at finite
temperature}}
\vspace{0.7cm}
\centerline{M.V. Cougo-Pinto \footnote{marcus@if.ufrj.br}, C. Farina
\footnote{farina@if.ufrj.br} and M. R. Negr\~ao \footnote
{guida@if.ufrj.br}}
\vspace{.5cm}
\centerline{\it  Instituto de F\'{\i}sica, Universidade Federal do Rio
de Janeiro }
\centerline{ CP 68528, Rio de Janeiro, RJ}
\centerline{ 21945-970, Brasil}
\vspace{.5cm}
\def\N{\rm l\ !N\,}
\def\Z{\angle \!\!\!{\rm Z}}

\begin{document}

\hspace\parindent
\begin{abstract}
\vspace{0.3cm}

We compute the combined effect of confinement, an external magnetic
field and temperature on the vacuum of the charged scalar field
using Schwinger's formula for the effective action in the imaginary
time formalism. The final result reproduces an effective Lagrangian
similar to the Heisenberg-Euler one in the limit of no confinement,
in the case of confinement it provides the necessary corrections to
this Lagrangian at each order of magnitude of the magnetic field.
The results show a finite temperature contribution to the vacum
permeability constant apart from the one due to confinement alone.

\end{abstract}
\vspace{.5cm}
\newpage

In a previous work \cite{promag0} we computed the combined effect
of confinement and an external magnetic field on the constitutive
relations of the vacuum of the charged scalar field. Knowing that
both the confinement and the applied electromagnetic field affect
the quantum vacuum of the charged scalar field \cite{promag0} it is
natural to ask what is the effect of considering the whole system
at finite temperature.

Let us consider the vacuum of the complex scalar field of mass $m$
and charge $e$ confined between two large parallel plates of side
$\ell$ and separation $a$, under the influence of an external
uniform constant magnetic field ${\bf B}$ with direction
perpendicular to the plates. The confinement is described by the
Dirichlet boundary conditions, which demand a vanishing field
within the plates. The system at temperature $1/\beta$ can be
described by its partition functiom $Z(\beta)$. From the
Schwinger's formulas for the partition function \cite{schwtemp} and
for the effective action at zero temperature\cite{schw51} it is
straightforward that they obey the relation:
\begin{eqnarray}
{\cal W}= i log Z(\beta).
\label{acaoez}
\end{eqnarray}

The expression for $logZ(\beta)$ \cite{casT} is given by:
\begin{equation}
\log{Z(\beta)}={1\over 2}\int_{s_o}^\infty
{ds\over s}\;Tr\,e^{-isH}\; ,
\label{Z}
\end{equation}
where $s_o$ is a cutoff in the proper-time $s$, $Tr$ means the
total trace and $H$ is the proper-time Hamiltonian in which the
frequencies have been discretized to the values $i2\pi n/\beta$
($n\in\Z$). For the charged scalar field we have
$H=(-i\partial-eA)^2+m^2$, where $e$ and $m$ are the charge and
mass of the field. We have for the trace in (\ref{Z}):
\begin{eqnarray}\label{Tr}
Tr\,e^{-isH}= 2e^{-ism^{2}}\sum_{n'=1}^{\infty}{eB\ell^2\over
2\pi}e^{-iseB(2n'+1)}
\sum_{n_{1}=1}^{\infty}e^{-is(\pi {n_{1}}/a)^2}
\sum_{n_{2}=-\infty}^{\infty}
e^{-is(2\pi n_{2}/\beta)^2} ,
\label{Tr}
\end{eqnarray}
where the first factor 2 is due to the charge multiplicity, the
first sum is on the Landau levels with the corresponding degeneracy
factor, the second sum is on the eigenvalues stemming from the
condition of confinement between the plates and the third sum is
due to the finite temperature condition. The sum on the Landau
levels is straightforward and the other two sums can be modified by
using Poisson summation formula \cite{poisson}:
\begin{eqnarray}
\sum_{n=-\infty}^{\infty} e^{-n^2\pi\tau}={1\over\sqrt{\tau}}
\sum_{n=-\infty}^{\infty} e^{-n^2\pi/\tau} .
\label{Poisson}
\end{eqnarray}
The trace is then given by:
\begin{eqnarray}
Tr\,e^{-isH}&=&
-\beta \frac{a\ell^2}{4{\pi}^{2}} \frac{e^{-ism^2}}{s^{2}}
[1 + iseB{\cal M}(iseB)]
\biggl [ \sum_{n_{1}=1}^\infty {e^{i(an_{1})^2/s}} -
\frac{\sqrt{i\pi s}}{2a} + \frac{1}{2} \biggl ]\times \nonumber\\
&\times&\biggl [ 1 +2\sum_{n_{2}=1}^\infty e^{i(\beta n_{2})^2/4s}
\biggl ],
\label{Trfinal}
\end{eqnarray}
where we have used the function ${\cal M}$ defined by ${\cal
M}(\xi)=cosech{\xi}-\xi^{-1}$. By using this expression for the
trace in (\ref{Z}) and substituting the result into
(\ref{acaoez})we get:
\begin{eqnarray}
-i\frac{\cal W}{a{\ell}^2 \beta}= L(\tilde B) + L(a, \tilde B) + L(\beta,
\tilde B) + L(a, \beta, \tilde B)
\label{acao1}
\end{eqnarray}
where
\begin{eqnarray}
L(\tilde B) = - \frac{1}{16{\pi}^{2}}
\int_{s_0}^{\infty} \frac{ds}{s^3}e^{-ism^2} [1 + (is{\tilde e}{\tilde
B}){\cal M}(is{\tilde e}{\tilde B})]
\label{HEB}
\end{eqnarray}
is the bosonic Euler-Heisenberg contribution to the effective
Lagrangian at zero temperature;
\begin{eqnarray}
L(a,\tilde B) &=&
\frac{i^{1/2}}{16{\pi}^{3/2}a}\int_{s_0}^{\infty}\frac{ds}{s^{5/2}}
e^{-ism^2}[1+(is{\tilde e}{\tilde B}){\cal M}(is{\tilde e}{\tilde
B})] +
\nonumber\\ &-& \frac{1}{8{\pi}^2} \sum_{n_1 =1}^{\infty}
\int_{s_0}^{\infty}\frac{ds}{s^3} e^{-ism^2 +i{(an_1)^2}/s}[1+(is{\tilde
e}{\tilde B}){\cal M}(is{\tilde e}{\tilde B})]
\label{HEC}
\end{eqnarray}
is the Euler-Heisenberg-Casimir contribution at zero temperature
\cite{promag0};
\begin{eqnarray}
L(\beta,\tilde B) = -\frac{1}{8{\pi}^2}\sum_{n_2 =1}^{\infty}
\int_{s_0}^{\infty} \frac{ds}{s^3} e^{-ism^2 + i{(\beta
{n_2}/2)^2}/s}[1+(is{\tilde e}{\tilde B}){\cal M}(is{\tilde
e}{\tilde B})]
\label{HEBT}
\end{eqnarray}
is the fintite temperature correction to the bosonic
Euler-Heisenberg effective Lagrangian and
\begin{eqnarray}
L(a,\beta,\tilde B)&=& -\frac{i^{1/2}}{8{\pi}^{3/2}a} \sum_{n_2
=1}^{\infty}
\int_{s_0}^{\infty} \frac{ds}{s^{5/2}} e^{-ism^2 + {i({\beta
n_2}/2)^2}/s}[1+(is{\tilde e}{\tilde B}){\cal M}(is{\tilde
e}{\tilde B})]+
\nonumber\\
&-& \frac{1}{4{\pi}^2} \sum_{n_1,n_2 =1}^{\infty}
\int_{s_0}^{\infty}\frac{ds}{s^3} e^{-ism^2+i[(an_1)^2 +({\beta
n_2}/2)^2]/s}\times \nonumber\\ &\times&[1+(is{\tilde e}{\tilde
B}){\cal M}(is{\tilde e}{\tilde B})]
\label{HECT}
\end{eqnarray}
is the finite temperature contribution to the
Euler-Heisenberg-Casimir effective Lagrangian.

The contributions given by equations (\ref{HEB}), (\ref{HEC}),
(\ref{HEBT}) and (\ref{HECT}) are unrenormalized and their
renormalization is required before remotion of the cutoff $s_{0}$.
An expansion of $1 + {\cal M}(iseB)$ in powers of $eB$ shows that
first term in $L(\tilde B)$ is a constant that can be subtracted
from the Lagrangian; in the limit $s_{0}
\rightarrow 0$ this constant tends to $m^{4} \Gamma (-2)/16 {\pi}^{2}$,
where $\Gamma$ is the Euler gamma function. The second term is
proportional to the Maxwell Lagrangian with a constant of
proportionality which tends to $e^{2} \Gamma (0)/48 {\pi}^{2}$ in
the limit $s_{0} \rightarrow 0$. This constant will be written as
$Z_{B}^{-1} -1$ and will be absorbed into Maxwell Lagrangian by
renormalization of $B$. The renormalized field will be defined as
$B_{R} = BZ_{B}^{-1/2}$ and the renormalized charge as $e_{R}
= eZ_{B}^{-1/2}$.  After subtractions and conversions to renormalized
quantities, $L(\tilde B)$ in (\ref{HEB}) becomes free of spurious
terms and well-defined in the limit $s_{o} \rightarrow 0$. The
Lagrangians $L(a,\tilde B)$, $L(\beta,\tilde B)$ and
$L(a,\beta,\tilde B)$ depend on $\tilde e$ and $\tilde B$ only
through the product $eB$ and we simply substitute on it $eB$ by
$\tilde e_{R}\tilde B_{R}$. The first two terms in the
 expansion of $1 + {\cal M}(iseB)$, which give spurious contributions to
(\ref{HEB}), give terms in (\ref{HEC}) and (\ref{HECT}) which
depend on the observable parameter $a$ and in the limit $s_{0}
\rightarrow 0$ are finite. As all quantities are properly
renormalized, the subindex $R$ will be omitted. So, the complete
effective Lagrangian is:
\begin{eqnarray}
{\cal L}&=& -\frac{1}{2}B^{2} + \frac{1}{16{\pi}^{2}}
\int_{0}^{\infty}
\frac{ds}{s^{3}} e^{-sm^{2}}[seB {\cal M}(seB) + \frac{(seB)^{2}}{6}]
+\nonumber\\ &+& \frac{1}{8{\pi}^{2}} \sum_{n_1 =1}^{\infty}
\int_{0}^{\infty}
\frac{ds}{s^{3}} \biggl [ e^{-ism^{2} + i(an_1)^{2}/s} +\frac{\sqrt{i\pi
s}}{2a} \biggl ][1 + (seB){\cal M}(seB)]\nonumber\\
&+&\frac{1}{8{\pi}^2} \sum_{n_{2}=1}^{\infty} \int_{0}^{\infty}
\frac{ds}{s^3} e^{-sm^{2}-{(\beta {n_{2}}/2)^2}/s}[1 + seB {\cal M}(seB)]
+\nonumber\\ &+&\frac{1}{4{\pi}^2} \sum_{n_{1},n_{2}=1}^{\infty}
\int_{0}^{\infty}
\frac{ds}{s^3} e^{-sm^{2}-[(an_{1})^2 + (\beta {n_{2}}/2)^2]/s}
[1 + seB {\cal M}(seB)]+\nonumber\\ &-&\frac{1}{8{\pi}^{3/2}a}
\sum_{n_{2}=1}^{\infty} \int_{0}^{\infty}
\frac{ds}{s^{5/2}} e^{-sm^{2}-{(\beta {n_{2}}/2)^2}/s}[1 + seB {\cal M}(seB)]
\label{lagtotal}
\end{eqnarray}
where the integration axis $s$ has been rotated to $-is$
\cite{schw51} and the cutoff $s_{0}$ has been finally removed.
There are five distinguished contributions in eq.(\ref{lagtotal}).
The first one is the renormalized Maxwell Lagrangian ${\cal
L}^{0}(B)$, given by the quadratic term $-B^{2}/2$. The second is
the renormalized version of the bosonic Heisenberg-Euler Lagrangian
(\ref{HEB})which can be expanded in powers of $B^2$ \cite{promag0}
to obtain:
\begin{eqnarray}
 L(B) = \sum_{k=2}^{\infty}
\frac{-m^{4}}{16{\pi}^{2}}\frac{2^{2k-1} - 1}{k(2k-1)(2k-2)}\biggl (
\frac{B^{2k}}{B_{cr}^{2k}} \biggl )^{2k} B_{2k}
\label{laghebr}
\end{eqnarray}
where $B_{2k}$ is the $2k$-th Bernoulli number. This expression
shows that the lowest order contribution from the bosonic
Heisenberg-Euler Lagrangian to Maxwell Lagrangian is a term
$B^{4}$. The third distiguished contribution to the complete
Lagrangian (\ref{lagtotal}) is given by the properly renormalized
version of the Lagrangian $L(a,B)$, which is written as
\cite{promag0}
\begin{eqnarray}
L(a,B)=\sum_{k=2}^{\infty} \frac{-m^{4}}{2{\pi}^{2}}
\frac{(2^{2k-1}-1)}{(2k)!} b_{2k}[\sum_{n \in \N}
(amn)^{2k-2}K_{2k-2}(2amn)] \frac{B^{2k}}{B_{cr}^{2k}}.
\label{BHEE}
\end{eqnarray}
This term provides the confinement correction to eq.
(\ref{laghebr}).

The fourth term in (\ref{lagtotal}), gives the finite temperature
correction to the Euler-Heisenberg free Lagrangian. After being
expanded in terms of $B^2$ it gives:
\begin{eqnarray}
L(\beta,B) = - \frac{m^2}{(\pi \beta)^2} \sum_{n_{2} =
1}^{\infty}\sum_{k=2}^{\infty}\frac{(-1)^{k-1}(2^{2k - 1} -
1)}{2^{2k-2}{\pi}^{2k}}\biggl (\frac{eB\beta}{2m}\biggl )^{2k}
n_{2}^{2k}\zeta(2k)K_{2k-2}(\beta m n_{2}).
\label{EHTEMP}
\end{eqnarray}
Finally, the fifth distinguished contribution to the complete
Lagrangian (\ref{lagtotal}) is given by the properly renormalized
finite temperature correction of the Euler-Heisenberg-Casimir
Lagrangian:
\begin{eqnarray}
L(a,\beta,B)&=&\frac{2m^4}{\pi^{2}}\sum_{n_1,n_2
=1}^{\infty}\frac{K_{2}(\sqrt{(2amn_{1})^2 + (\beta mn_{2})^2})}
{(2amn_{1})^2 + (\beta mn_{2})^2} +\nonumber\\ &-&
\frac{1}{2}\biggl [1-\frac{1}{\mu (am,\beta m)}\biggl ]B^2 +
L'(a,\beta,B)
\label{HECTemp}
\end{eqnarray}
where
\begin{eqnarray}
\frac{1}{\mu (am,\beta m)}&=&  1+ \frac{e^2}{12\pi^2} \biggl [
\sum_{n_1=1}^{\infty} K_0(2amn_1) +
\sum_{n_2=1}^{\infty} K_0(\beta mn_2) +\nonumber\\
&+&\sum_{n_1,n_2 =1}^{\infty} K_0(\sqrt{(2amn_1)^2 + (\beta
mn_2)^2})\biggl ]+\nonumber\\ &+& \frac{e^2}{48\pi am} [1+
\sum_{n_2 =1}^{\infty} e^{-\beta mn_2}]
\label{permT}
\end{eqnarray}
and
\begin{eqnarray}
L'(a,\beta,B) &=& \frac{1}{4\pi^2}\sum_{n_1,n_2 =1}^{\infty}
\int_{0}^{\infty} \frac{ds}{s^3}e^{-sm^2-[(an_1)^2+({\beta n_2}/2)^2]/s}
\biggl [ seB {\cal M}(seB) + \frac{(seB)^2}{6} \biggl ] +\nonumber\\
&-& \frac{1}{8\pi^{3/2}a} \sum_{n_2 =1}^{\infty} \int_{0}^{\infty}
\frac{ds}{s^{5/2}}e^{-sm^2 -{(n_2 \beta/2)^2}/s} \biggl [seB{\cal M}(seB)
+ \frac{(seB)^2}{6} \biggl ]
\label{Llinha}
\end{eqnarray}
The first term in (\ref{HECTemp}) is minus the Casimir energy
density for a massive bosonic field (cf. \cite{mfn1} and references
therein) at finite temperature. It is an observable quantity, but
it has no importance in the present formalism because it is
independent of $B$. If our sole interest is to obtain an effective
Lagrangian for the magnetic field $B$ at finite temperature, this
first term can be simply discarded. The term $L'(a,\beta,B)$ in
(\ref{HECTemp}), which is given by (\ref{Llinha}), is the
correction due to finite temperature to the
Casimir-Heisenberg-Euler Lagrangian $L(a,B)$. The contribution
(\ref{Llinha}) can be expanded in terms of $B^{2}$ to give (cf
formula 3.471,9 in \cite{gradsh}):
\begin{eqnarray}
L'(a,\beta ,B)&=& - \frac{1}{2\pi^2}\sum_{n_1,n_2 =1}^{\infty}
\sum_{k=2}^{\infty}\frac{(-1)^{k-1}(2^{2k-1}-1)}{2^{2k-2}\pi^{2k}}
\zeta(2k)(eB)^{2k}\times \nonumber\\
&\times& \biggl [\frac{(2an_{1})^2 + (\beta n_{2})^2}{(2m)^2}
\biggl ]^{k-1}K_{2k-2}(\sqrt{(2amn_{1})^2 + (\beta
mn_{2})^2})+\nonumber\\ &+&\frac{1}{4\pi^{3/2}a}\sum_{n_2
=1}^{\infty}
\sum_{k=2}^{\infty}\frac{(-1)^{k-1}(2^{2k-1}-1)}{2^{2k-2}\pi^{2k}}
\zeta(2k)(eB)^{2k}\biggl (\frac{\beta n_{2}}{2m}
\biggl )^{2k-3/2} \times \nonumber\\
&\times&K_{2k-3/2}(\beta mn_{2})
\end{eqnarray}
This expression provides a term by term correction to the
expression (\ref{laghebr}) of the bosonic Euler-Heisenberg
effective Lagrangian in powers of $B^2$.

The second term in (\ref{HECTemp}), which is quadratic in the field
$B$, provides a contribution to the effective Lagrangian that is
totally absent in the bosonic Euler-Heisenberg Lagrangian. This
term is the finite temperature correction to the confinement
contribution found in \cite{promag0}. It can be expressed as a
change in the vacuum permeability constant due to both confinement
and temperature corrections and is given by (\ref{permT}). To
highlight its importance let us consider a weak field regime in
which only quadratic terms in $B$ are not negligible. In this
situation the bosonic Heisenberg-Euler Lagrangian (\ref{laghebr}),
as well as its corrections in (\ref{BHEE}), (\ref{EHTEMP}) and
(\ref{Llinha}), makes no contribution to the effective Lagrangian of
the field $B$. The only contribution that is left is the quadratic
one from (\ref{HECTemp}), which changes the Lagrangian
(\ref{lagtotal}) to
\begin{eqnarray}
{\cal L} = - \frac{1}{2}\frac{B^{2}}{{\mu}(am,\beta m)}.
\label{maxlagcor}
\end{eqnarray}

In the general case the permeability $\mu (am,\beta m)$ of the
bosonic confined vacuum, which is given by
\begin{eqnarray}
\mu (am,\beta m)&=& \{ 1+ \frac{e^2}{12\pi^2} \biggl [
\sum_{n_1=1}^{\infty} K_0(2amn_1) +
\sum_{n_2=1}^{\infty} K_0(\beta mn_2) +\nonumber\\
&+&\sum_{n_1,n_2 =1}^{\infty} K_0(\sqrt{(2amn_1)^2 + (\beta
mn_2)^2})\biggl ]+\nonumber\\
&+& \frac{e^2}{48\pi am} [1+
\sum_{n_2 =1}^{\infty} e^{-\beta mn_2}] \}^{-1}
\label{perme}
\end{eqnarray}
reduces to 1 when both confinement and temperature effects
disappears, ($a \rightarrow \infty$) and ($\beta \rightarrow
\infty$). From the properties of the Bessel functions \cite{gradsh}
it is easy to see that it decreases quickly from infinity to zero
as $am$ and $\beta m$ increase from zero to infinity, so their
effect on the confinement is a very small reduction on the magnetic
permeability. The other term gives a more significant contribution
to the expression. The above expression gives the finite
temperature correction for the magnetic permeability behaviour of
the confined bosonic vacuum, which showed clearly a diamagnetic
structure at zero temperature.

\end{document}